\documentclass[10pt,conference]{IEEEtran}

\usepackage{amsfonts}
\usepackage{setspace}

\usepackage{setspace}
\usepackage{amssymb}
\usepackage{epsfig}
\usepackage{graphicx}
\usepackage{xcolor}
\usepackage{algorithm}
\usepackage{algorithmicx}
\usepackage{algpseudocode}
\usepackage{amsmath}
\usepackage{cite}

\floatname{algorithm}{Algorithm}

\usepackage[ps2pdf,
bookmarks=false,
bookmarksnumbered=false, 
bookmarksopen=false, 
colorlinks=false]{}

\begin{document}

\title{Deep Multi-Agent Reinforcement Learning Based Cooperative Edge Caching in Wireless Networks}

\author{Chen Zhong, M. Cenk Gursoy, and Senem Velipasalar
\\Department of Electrical Engineering and Computer Science,
Syracuse University, Syracuse, NY 13244
\\Email: czhong03@syr.edu, mcgursoy@syr.edu, svelipas@syr.edu}

\maketitle

\begin{abstract}
	The growing demand on high-quality and low-latency multimedia services has led to much interest in edge caching techniques. Motivated by this, we in this paper consider edge caching at the base stations with unknown content popularity distributions. To solve the dynamic control problem of making caching decisions, we propose a deep actor-critic reinforcement learning based multi-agent framework with the aim to minimize the overall average transmission delay. To evaluate the proposed framework, we compare the learning-based performance with three other caching policies, namely least recently used (LRU), least frequently used (LFU), and first-in-first-out (FIFO) policies. Through simulation results, performance improvements of the proposed framework over these three caching algorithms have been identified and its superior ability to adapt to varying environments is demonstrated.

\end{abstract}


\section{Introduction}

According to \cite{indexglobal}, the mobile data traffic has grown 17-fold from 2012 to 2017, and $59 \%$ of the global mobile data traffic is generated by the demand for videos in 2017. In addition, a 9-fold increase in video data traffic is predicted by 2022, making the mobile video traffic to account for $79\%$ of total mobile traffic. In contrast, the three-fold increase in the  mobile network connection speed is expected to be inadequate for satisfying users' demands on high-quality multimedia streaming services.

As a promising technique to reduce the congestion in data traffic, content caching has received considerable attention in recent years. Indeed, caching-based content distribution is employed in content delivery networks (CDNs). CDN is widely adopted to reduce the data congestion near the content server. Usually, to solve the CDN in-network caching problem, routing policy to allocate users' requests to different servers needs to be addressed. Prior work \cite{dehghan2017complexity, xu2018joint} has focused on jointly solving the caching and routing problems to minimize the service delay. Though the application of CDN has been shown to reduce the data traffic, it can hardly handle the growing mobile data traffic because it is inevitable that the content has to be transmitted through the CDN nodes before arriving at the user. More recently, proactive caching at the wireless network edge, such as at the base stations and user equipments, is proposed. This technique makes it possible to have popular contents to be placed closer to the end users and be directly transmitted, which can effectively reduce the time for routing in CDNs, and apparently save a considerable amount of waiting time for users and offload a portion of data traffic at the CDNs. For instance, authors in \cite{zhang2018cooperative} and \cite{li2018learning} studied edge caching policies aimed at minimizing the transmission delay for the base station and D2D users.

In addition to the decision on where to locate the cache, the study of cache replacement policies is also of great importance. In the literature, different methods have been applied to determine optimal caching policies. For the case of decentralized caching, the authors presented in \cite{kvaternik2016methodology} a decentralized optimization method for the design of caching strategies that aimed at minimizing the energy consumption of the network, while in \cite{wang2018decentralized}, a decentralized framework for proactive caching is proposed based on blockchains considering a game-theoretic point of view. In \cite{zhou2017optimal}, caching and multicast problems are jointly solved using dynamic programing. Moreover, the machine learning techniques are also applied in this field. In \cite{lei2017deep}, deep neural networks are used to train the caching optimization algorithms. And to better adapt to changing environments, the application of reinforcement learning algorithms is proposed. For example, a deep actor-critic framework for content caching at the base station is proposed in \cite{zhong2018deep}. And to seek for the optimal cooperative caching policy in a decentralized caching network, the authors in \cite{song2017learning} presented a multi-armed bandit based solution, while the authors in \cite{sung2016efficient} adopted a multi-agent Q-learning scheme.

In this paper, we also investigate the edge caching problem at the base stations. For a system with multiple cache-enabled base stations, if the system is aimed at achieving the overall minimum transmission delay, the relationship between base stations is both competitive and cooperative. Motivated by the performance of multi-agent framework presented in \cite{lowe2017multi, foerster2016learning, gupta2017cooperative}, we propose a deep multi-agent actor-critic framework based on a partially observable Markov decision process (POMDP) to solve this decentralized caching problem.

Our main contributions are listed as follows:
\begin{itemize}
	\item We propose a multi-agent actor-critic framework for content caching problem. We describe the observations and actions for each agent, and employ this multi-agent system to make caching replacement decisions to minimize the transmission delay.
	\item The proposed framework operates without having any information on content popularities and user preferences.
	\item The proposed framework can effectively help to reduce the transmission delay and adapt to the changing environments.
	\item We demonstrate the performance improvements of this proposed framework over other caching policies.
\end{itemize}

\section{System Model}
\subsection{System Model}

\begin{figure}
\centering
\includegraphics[width=1.0\linewidth]{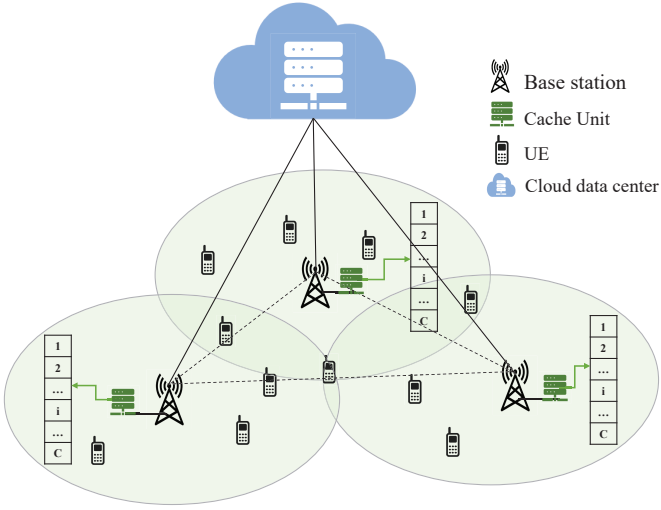}
\caption{System Model}
\label{fig:system-model}
\end{figure}

As shown in Fig. \ref{fig:system-model}, we consider a communication network with a cloud data center and $N$ base stations. It is assumed that the data center has sufficient storage space to cache all content files, while each base station has a fixed cache capacity of $C$. All base stations can connect with the cloud data center and request files from it. And the base station can decide whether to cache the file or not. Each base station covers a fixed cellular region described by a circle with the corresponding base station at the center. We assume that the radii of the cells are fixed and all users in the cell can access the corresponding base station. There are $U$ users randomly distributed in the system, and they are located in at least one cellular region covered by a base station to ensure service. We assume that in a given time slot, the users' locations do not change and those located at the overlapped regions can be served by any one of the corresponding base stations. Users have their own preferences for contents, and in each time slot each user can request only one content. Here, we denote the total number of contents as $M$, and use the content ID to denote the requests for the corresponding content. In each operation cycle, users request a content based on their own preferences. The requests are sent to all base stations that can connect with the user, and the base station that can provide minimum transmission delay will finally transmit the requested content file to the user. In the meantime, all base stations will update their caches to minimize the average transmission delay based on the users' requests.


The base stations will compete with each other to get the chance to transmit and also cooperate with each other to reduce the overall transmission delay. To realize this framework, we proposed an actor-critic reinforcement learning based multi-agent framework. In this framework, there are $N$ actor networks and one centralized critic network. We consider each base station as an agent that adopts one of the actor networks to seek its own caching policy.
And we assume that there are control channels that allow the base stations to send the caching state and data traffic parameters to the cloud date center, so that the cloud data center can act as the centralized critic to evaluate the overall caching state. Similarly as in \cite{zhong2018deep}, in each operation cycle, the agent can either keep the cache state the same or replace unpopular contents with the popular ones. Note that there can be more than one request arriving at a base station at the same time, and for different contents, the agent needs to jointly decide which cached content will be deleted and which content requested by which user will be cached. We define the action space as $\mathcal{A}$, and let $\mathcal{A} = \left\lbrace a_0, a_1,..., a_{\mathcal{D}}\right\rbrace $, where $a_{\nu}$ denotes a valid action. In our case, $a_0$ indicates that the current cache state is unchanged. For $\nu = \{1, 2, ..., \mathcal{D}\}$, we define $\mathcal{D} = \binom{\mathcal{C}_i}{1} \binom{L_i}{1} $, where $\mathcal{C}_i$ is the number of files in the cache of base station $i$, and $L_i$ is the number of users that can connect with the base station $i$. So each $a_\nu$ stands for a possible combination to replace one of $\mathcal{C}_i$ cached contents with one of $L_i$ currently requested contents. For every time slot, all agents must select their own action from the action space $\mathcal{A}$ and execute.

\subsection{Transmission Delay}\label{Sec:delay}
In this work, we formulate the performance of the caching policy in terms of transmission delay. The transmission delay is defined as the number of time frames needed to transmit a content file, and can be expressed as

\begin{align}
T=\min\left\{\tilde{t}:F\leq\sum_{\kappa=1}^{\tilde{t}} T_0 C[\kappa]\right\} \label{eq:transmissiondelay}
\end{align}
where $F$ is the size of the content file to be transmitted. $T_0$ stands for the duration of each time frame, and $C[\kappa]$ is the instantaneous channel capacity in the $\kappa^{\text{th}}$ time frame. And the channel capacity $C[\kappa]$ is expressed as

\begin{align}
C[\kappa]=B\log_2\left(1+\frac{P_t}{B\sigma^2}z_\kappa\right)\hspace{0.4cm} \text{bits}/\text{s}
\end{align}
where $P_t$ is the transmission power, $B$ is the bandwidth, $\sigma^2$ is the noise power spectral density, and $z_\kappa$ is the magnitude square of the corresponding fading coefficient in the $\kappa^{\text{th}}$ time frame. In the system, there are two types of transmitters: the cloud data center and the base stations. We assume that all transmitters transmit at their maximum power level to maximize the transmission rate. The transmission power is defined as

\begin{align}
P_t=
\begin{cases}
P_b\hspace{1cm} \text{if the transmitter is the cloud data center}\\
P_i\hspace{1cm} \text{if the transmitter is the $i^{\text{th}}$ base station}
\end{cases}.
\end{align}
So, if user $j$ requests a content, which is not cached at any base station that can connect with the user, the content file will be first transmitted from the cloud data center to the base station $\hat{i}$, which is the closest base station to the user $j$, and then from the base station $\hat{i}$ to user $j$. Thus, the minimum transmission delay $\hat{D}_j$ in the case of missing file in the cache can be expressed as

\begin{equation}
\hat{D}_j = T_{c,\hat{i}} + T_{\hat{i}, j}
\end{equation}
where $T_{c,\hat{i}}$ stands for the transmission delay form the cloud data center to the base station $\hat{i}$, and $ T_{\hat{i}, j}$ is the transmission delay from the base station $\hat{i}$ to the user $j$.

However, if the requested file is cached at a base station $i$, which can connect to user $j$, the transmission delay $D_j$ for the case of hitting the cache can be expressed as
\begin{equation}
D_j = T_{i, j}
\end{equation}

\section{Problem Formulation}
In the previous section, we have described the transmission delay for both cases of missing and hitting the cache. In this section, we formulate the caching problem. Firstly, we define the transmission delay reduction $\Delta D_j$ as
\begin{equation}
\Delta D_j = \hat{D}_j - D_j.
\end{equation}

Now, the average transmission reduction in an operation cycle is
\begin{align}
\Delta D &= \frac{1}{U} \sum\limits_{j = 1}^{U} \Delta D_j\\
& = \frac{1}{U} \sum\limits_{j = 1}^{U}(\hat{D}_j - D_j )\\
& = \frac{1}{U} \sum\limits_{j = 1}^{U} (T_{c,\hat{i}} + T_{\hat{i}, j} - T_{i, j} )
\end{align}
where $U$ is the total number of users. In this work, our goal is to maximize the average transmission delay reduction, and the caching problem is formulated as follows:

\begin{align}
\textbf{P1:}\hspace{1cm}&\underset{\mathbf{\Phi}}{\text{Maximize}}\hspace{1.2cm} \Delta D\\
&\text{Subject to} \hspace{0.7cm} \xi_{i,j} = 1 \quad \exists i \,\, \forall j \\
&\hspace{2.1cm} \sum_{f = 1}^{M} \phi_{i, f} F_f \leq C \label{condition:12}
\end{align}
where $\mathbf{\Phi}$ is an $N \times M$ matrix which records the caching states of the $N$ base stations, and each element $\phi_{i, f}$ in the caching state matrix is an indicator to show if the file is cached:

\begin{align}
\phi_{i, f}=
\begin{cases}
1\hspace{0.4cm} \text{if the file $f$ is cached at the base station $i$}\\
0\hspace{0.4cm} \text{if the file $f$ is not cached at the base station $i$}
\end{cases}\hspace{-.3cm}.
\end{align}
$F_f$ is the size of file $f$. Since in this work we assume all files have the same size, the condition in (\ref{condition:12}) can be rewritten as
\begin{equation}
	 \sum_{f = 1}^{M} \phi_{i, f} \leq \mathcal{C}
\end{equation}
where $\mathcal{C}$ is the maximum number of files that can be stored at each base station.
And $\xi_{i,j}$ is an indicator describing if user $j$ is in the area covered by base station $i$:
\begin{align}
\xi_{i,j}=
\begin{cases}
1\hspace{0.2cm} \text{if user $j$ can connect to base station $i$}\\
0\hspace{0.2cm} \text{if user $j$ cannot connect to base station $i$}
\end{cases}.
\end{align}

\section{Multi-agent Actor-Critic Content Caching Framework}\label{sec:framework}

In this section, we present the multi-agent actor-critic content caching framework. Since each user has a unique preference, the requests arriving at different base stations are also very different, and this requires each base station to run a unique caching policy based on the preferences of the users in the base station's service range. However, considering the users in the overlapped areas, if the system is aimed to achieve an overall high performance, it is unavoidable that base stations need to communicate and cooperate. With this purpose, we introduce a multi-agent framework with centralized critic and decentralized actor, and consider every base station as an agent.

\subsection{Multi-Agent Actor-Critic}\label{sub:algorithm}
We introduce a multi-agent actor critic framework based on the partially observable Markov decision processes with $N$ agents, where the critic network $V(\mathtt{x})$ and $N$ actors $\pi_{\theta_i}(o_i)$, $i = i, 2, ..., N$, are parameterized by $\theta = \{\theta_c, \theta_1, \theta_2, ..., \theta_N \}$.

\emph{Actor:} The actor network is defined as a function to seek a caching policy $\pi = \{\pi_1, \pi_2, ..., \pi_N \}$, which can map the observation of the agent to a valid action chosen from the action space $\mathcal{A}$. In each time slot, agent $i$ will select an action $a_i$ based on its own observation $o_{i}$ and policy $\pi_i$:
\begin{equation}
a_i = \pi_i(o_i).
\end{equation}


\emph{Critic:} The critic is employed to estimate the value function $V(\mathtt{x})$, where $\mathtt{x}$ stands for the observation of all agents, $\mathtt{x} = \{o_1, o_2, ..., o_N \}$. At time instant $t$, after the actions $a_t = \{ a_{1,t}, ..., a_{N,t} \}$ are chosen by the actor networks, the agents will execute the actions in the environment and send the current observation $\mathtt{x}_t$ along with the feedback from the environment to the critic. The feedback includes the reward $r_t$ and the next time instant observation $\mathtt{x}_{t+1}$. Then, the critic can calculate the TD (Temporal Difference) error:
\begin{equation}
\delta^{\pi_\theta} = r_t + \gamma V(\mathtt{x}_{t+1}) - V(\mathtt{x}_t)
\end{equation}
where $\gamma \in (0,1)$ is the discount factor.

\emph{Update:} The critic is updated by minimizing the least squares temporal difference (LSTD):
\begin{equation}
V^* = \arg \min_{V} (\delta^{\pi_\theta} )^2
\end{equation}
where $V^*$ denotes the optimal value function.

The actor $i$ is updated by policy gradient. Here we use TD error to compute the policy gradient:
\begin{equation}
\nabla_{\theta_{i}} J(\theta_i) = E_{\pi_{\theta_i} } [ \nabla_{\theta_i} \log \pi_{\theta_i}(o_i, a_{i})  \delta^{\pi_{\theta}} ]
\end{equation}
where $\pi_{\theta_i}(o_i, a_i)$ denotes the score of action $a_i$ under the current policy. Then the weighted difference of parameters in the actor $i$ can be denoted as $\Delta\theta_{i} = \alpha \nabla_{\theta_{i}} \log \pi_{\theta_i}(o_{i}, a_{i}) \delta^{\pi_\theta}$, where $\alpha \in (0,1)$ is the learning rate. And the actor network $i$ can be updated using the gradient decent method:
\begin{equation}
\theta_i \longleftarrow \theta_i + \alpha \nabla_{\theta_{i}} \log \pi_{\theta_i}(o_{i}, a_{i}) \delta^{\pi_\theta} 
\end{equation}

The complete algorithm is shown below in Algorithm \ref{alg:multiagent}.
\\
\\
\\

\begin{algorithm}[H]
	\caption{Multi-Agent Actor-Critic Algorithm for Content Caching}
	\label{alg:multiagent}
	\begin{algorithmic}
		\State Initialize critic network $V(\mathtt{x})$ and actor $\pi_{\theta_i}(o_i)$, parameterized by $\theta = \{\theta_c, \theta_1, \theta_2, ..., \theta_N  \}$.
		\State Receive initial state $\mathtt{x} = \{o_1, o_2, ..., o_N\}$.		
		\For{$t = 1,T$}
		\State The base station receives users' requests $Req_t = \{req_{1,t}, req_{2,t}, ..., req_{U,t}\}$.
		\State Extract observation at time $t$ for each agent, and $\mathtt{x}_t = \{o_{1,t}, o_{2,t}, ..., o_{N,t}\}$.
		\State For each agent $i$, select action $a_i = \pi_{\theta_i}(o_{i,t} )$ w.r.t. the current policy.
		\State Execute actions $a_t = (a_{1,t}, a_{2,t}, ..., a_{N,t})$ to update the cache state of each base station.
		\State Observe reward $r_t$ and new state $\mathtt{x}_{t+1}$.
		\State Critic calculates the TD error based on the current parameter: $ \delta^{\pi_\theta} = r_t + \gamma V(\mathtt{x}_{t+1}) - V(\mathtt{x}_{t}) $.
		\State Update the critic parameter $\theta_c$ by minimizing the loss: $\mathcal{L}(\theta) = (\delta^{\pi_\theta} )^2$.
		\For{agent $i = 1$ to $N$}
		\State Update the actor policy by maximizing the action value: $\Delta\theta_{i} = \alpha \nabla_{\theta_{i}} \log \pi_{\theta_i}(o_{i,t}, a_{i}) \delta^{\pi_\theta}$, $\alpha \in (0,1)$.
		\EndFor	
		\State Update features space $\mathcal{F}$.
		
		\EndFor
	\end{algorithmic}
\end{algorithm}

\subsection{Environment}
To perform the experiments, we consider a wireless cellular network with $N$  base stations with specific service ranges as shown in Figure \ref{fig:b5u30}. Allowing the agents to make their own caching decisions and cooperate with each other, the framework is proposed as a centralized critic network together with a decentralized actor network. Therefore, the agents will feed the actor network with their own observations and feed the critic network with the complete state space.

\emph{Agents' Observation and State Space:}
As introduced in previous sections, the multi-agent actor critic framework is based on a partially observable Markov decision process. Each agent $i$ for $i = 1, 2, ..., N$, can only observe the requests arriving at itself, and select its own action only based on the observation $o_i$. In the environment, agent $i$ can observe the contents' features through its local request history. For the centralized critic, the state space is defined as $\mathtt{x} = \{o_1, o_2, ..., o_N \}$.

\emph{Feature Space:}
The feature space consists of three components: short-term feature $\mathcal{F}_s$, medium-term feature $\mathcal{F}_m$, and long-term feature $\mathcal{F}_l$, which represent the total number of requests for each content in a specific short-, medium-, long-term, respectively. These features are updated as the new requests arrive at agents. Then, we let $f_{xj}$, for $x\in \left\lbrace s, m, l\right\rbrace $ and $j\in\{1,\ldots, M\} $, denote the feature of a specific content within a specific term, where $M$ is the total number of contents. Thus, the observation for each agent $i$ is defined as $o_i = \{\mathcal{F}_s; \mathcal{F}_m; \mathcal{F}_l\}$ where
$\mathcal{F}_s=\{f_{s0}, f_{s1},..., f_{sM}\}$,
$\mathcal{F}_m=\{f_{m0}, f_{m1},..., f_{mM}\}$, and
$\mathcal{F}_l=\{f_{l0}, f_{l1},..., f_{lM}\}$.

\emph{Reward:}
In this work, we consider the objective function in problem $\textbf{P1}$ as the reward. In every operation cycle $t$, after the agents update their caches according to the selected actions, the average delay reduction of transmitting the content files requested by the users in the next operation cycle $t+1$ will be received as the reward within the multi-agent framework. So we define the reward in the $t^{\text{th}}$ operation cycle as

\begin{equation}
r_t = \Delta D^{t+1}
\end{equation}

\section{Simulation Results}

In this section, we demonstrate the simulation results. To better evaluate the proposed framework, we compare its performance with the following caching algorithms:
\begin{itemize}
	\item \textbf{Least Recently Used (LRU) :} In this policy, the system keeps track of the most recent requests for every cached content. And when the cache storage is full, the cached content, which is requested least recently, will be replaced by the new content.
	\item \textbf{Least Frequently Used (LFU) :} In this policy, the system keeps track of the number of requests for every cached content. And when the cache storage is full, the cached content, which is requested the least many times, will be replaced by the new content.
	\item \textbf{First In First Out (FIFO) :} In this policy, the system, for each cached content, records the time when the content is cached. And when the cache storage is full, the cached content, which was stored the earliest, will be replaced by the new content.
\end{itemize}

In their implementation, the above three caching policies are executed at each base station independently.

\subsection{Simulation Setup}

\begin{figure}
	\centering
	 \includegraphics[width=0.9\linewidth]{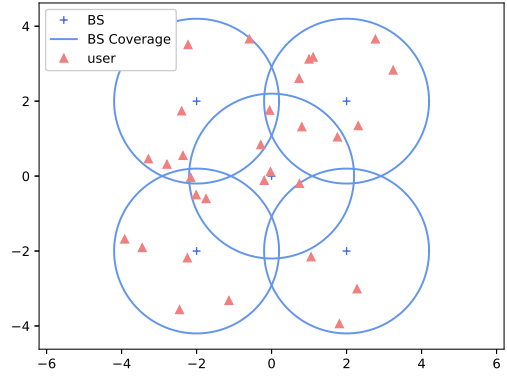}
	\caption{Coverage map of a system contains 5 base stations and 30 users}
	\label{fig:b5u30}
\end{figure}

\emph{Environment Settings:} As shown in Fig. \ref{fig:b5u30}, in the experiments, we consider a system with 5 base stations and 30 users randomly distributed in the area, each covered by at least one of the base stations. The cell radius is set as $R = 2.2$km, and the transmission power of all base stations is set as $P_i = 16.9$dB, $i = 1,2, ..., 5$. The transmission power of the cloud data center is set as $P_c = 20$dB. As assumed, the content files are split into units of the same size, and the size of each unit is set as $96.13$bits. And we assume Rayleigh fading with path loss $\mathbb{E}\{z\} = d^{-4}$, where $d$ is the distance between the transmitter and receiver.

\emph{File/Content Request Generation:} In our simulations, the raw data of users' requests is generated according to the Zipf distribution
\begin{equation}
f(k; \beta, M) = \frac{1/k^{\beta}}{\sum_{m = 1}^{M}(1/m^{\beta})}
\end{equation}
where the total number of files $M$ is set as 500, and the Zipf exponent $\beta$ is fixed at 1.3 in the study of the cache size, while it  varies when the impact of the Zipf parameter is considered. $k$ is the rank of the file, and in the implementation, a user's preference for files is randomly generated. To encourage the base station to cache the files that are popular for more users, the users are randomly divided into 5 groups. It is assumed that the users in the same group will have similar but not exactly the same rank for all files. And the group information will not influence the users' location. It is important to note that while we generate the requests with Zipf distribution and also group the users, such information is totally unknown to the agents.

\emph{Feature Extraction:} From the raw data of content requests, we extract the feature $F$ and use it as the agents' observations of the network. Here, as features, we consider the number of requests for a file within the most recent $10$, $100$, $1000$ requests.

\subsection{Transmission Delay}
In this section, we present the the simulation results. We evaluate the reduction in transmission delay as a percentage as follows:
\begin{equation}
\eta = \frac{\Delta D}{\frac{1}{U}\sum\limits_{j = 1}^{U} \hat{D}_j} \times 100\%
\end{equation}
Hence, $\eta$ is the percentage of delay reduction per user in one operation cycle.

\begin{figure}
\centering
\includegraphics[width=1.0\linewidth]{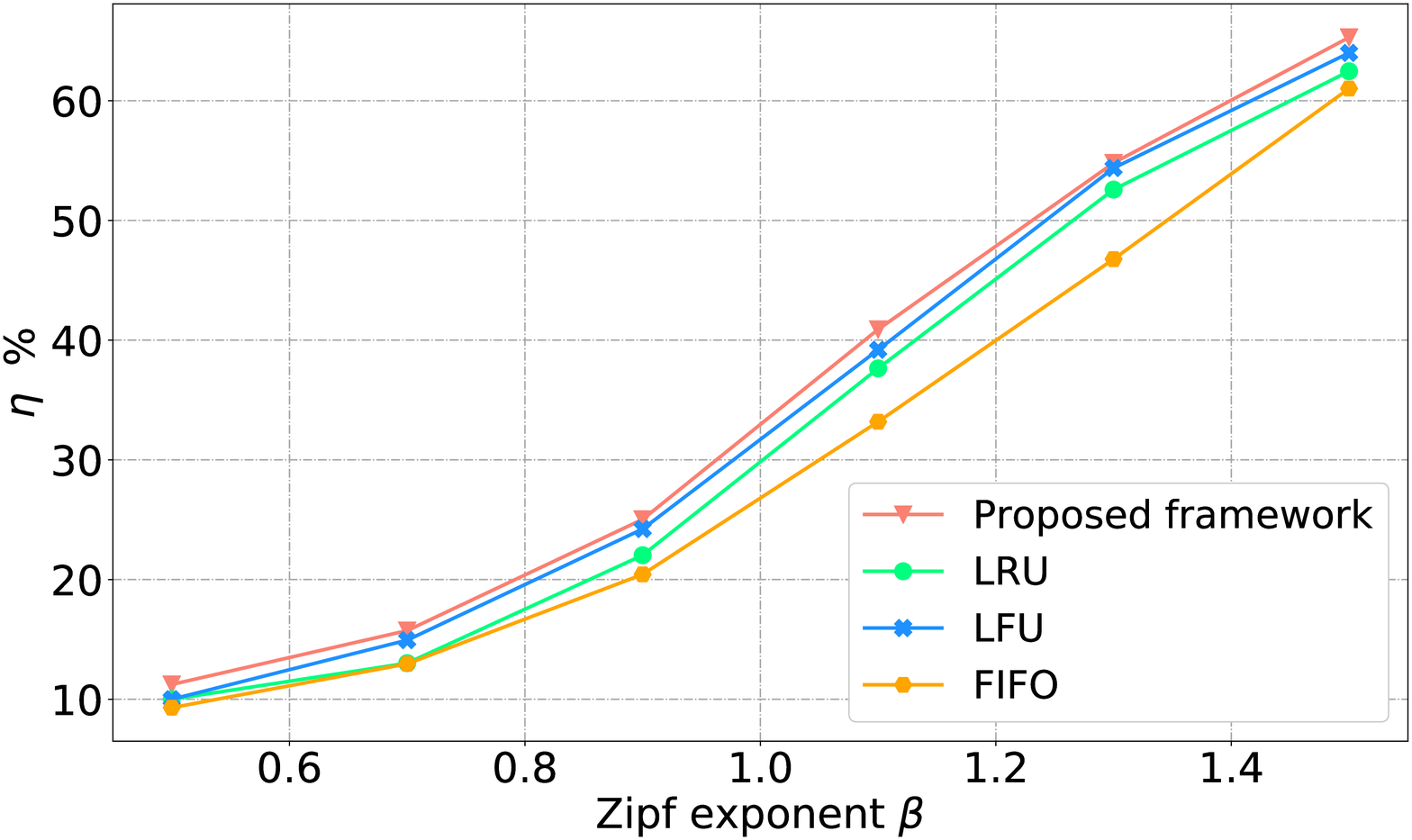}
\caption{Percentage of transmission delay reduction $\eta$ vs. Zipf exponent $\beta$}
\label{fig:beta}
\end{figure}

In Fig. \ref{fig:beta}, we fix the maximum number of content files that can be cached at the base station as $\mathcal{C} = 40$, and plot the percentage of overall average transmission delay reduction $\eta$ as a function of the Zipf exponent, which varies as $\{0.5, 0.7, 0.9, 1.1, 1.3, 1.5 \}$. We can observe that the proposed framework shows improvements for all values of $\beta$. Note that LRU and LFU can also choose contents that are relatively popular to cache. Note also that the value of $\beta$ reflects the popularity distribution of contents. As $\beta$ decreases, the distribution of the content popularity tends to become uniform, indicating that the difference in the popularities of contents with higher and lower ranks is smaller. This makes it more difficult for the caching policy to find the most popular files via learning the statistics of the users' requests. The observation that our proposed framework leads to improved performance when the value of $\beta$ is small means that the proposed framework is more competitive in coping with the disturbance from the less popular contents, which makes it more suitable to be applied in the case in which the popularity pattern is less distinguishable.

\begin{figure}
\centering
\includegraphics[width=1.0\linewidth]{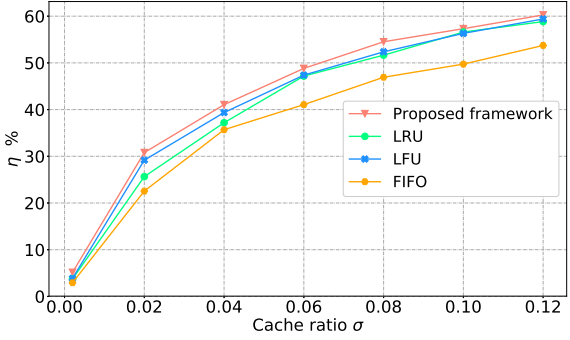}
\caption{Percentage of transmission delay reduction $\eta$ vs. cache ratio $\sigma$}
\label{fig:cachesize}
\end{figure}

To determine the relationship between transmission delay and cache capacity, in Fig. \ref{fig:cachesize}, we fix the Zipf exponent at $\beta = 1.3$, and plot the percentage of overall transmission delay reduction $\eta$ as a function of the cache capacity. However, instead of directly using the cache capacity $C$, we consider the cache ratio $\sigma = \frac{\mathcal{C}}{M}$ (where $M$ is the total number of content files that can be requested by the users), so that we can analyze the impact of the cache capacity normalized by the potential data traffic flows into this system. It is shown that as the cache ratio $\sigma$ increases, the reduction in transmission delay achieved by all four caching policies first rises quickly because the base stations can cache more files, and then the trend slows down after a certain value of $\sigma$. The upward trend starts to slow down because all these caching algorithms are encouraged to cache the most popular files following the statistics they learn. So when the cache ratio grows further and further, the caching agent will start caching the less popular content files. Though more files are cached and transmission delay is further reduced, caching the less popular files at the edge nodes lead to smaller improvements in reducing the transmission delay when compared with the contribution made by caching the most popular files. In an other words, when the cache ratio is large enough to cache all of the most popular files, the system does not necessarily have to keep enlarging the cache capacity, considering the price to pay for the storage and the relatively small reduction in transmission delay that will be achieved by storing the less popular files. We also observe that for all values of the cache ratio, the proposed framework achieves better performance for two reasons: First, the proposed framework considers the reduction in the average transmission delay as the reward, so that the caching algorithm does not only focus on finding the most popular files, but takes into account the users' locations and several less popular files with potentially high delay penalties if not cached; and secondly, the critic network can facilitate the exchange of information among the base stations so that they can avoid caching the same file to serve the user located in the overlapped regions, and in this way, utilize the cache space more efficiently.

\begin{figure}
\centering
\includegraphics[width=1.0\linewidth]{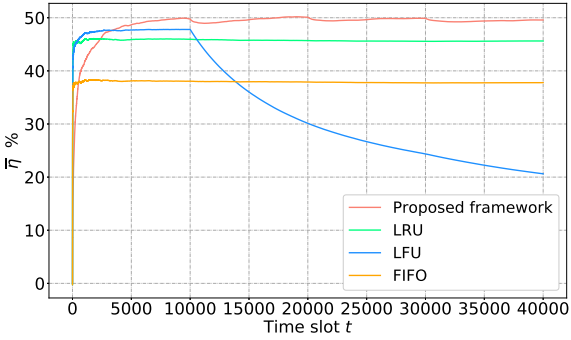}
\caption{Percentage of transmission delay reduction $\eta$ as the popularity distribution of contents change over time}
\label{fig:timeVarying}
\end{figure}

In Fig. \ref{fig:timeVarying}, we demonstrate the ability of the caching policies to adapt to varying content popularity distributions. In this experiment, the users' preferences for files change at every $10000$ time slots. The users' requests are generated using Zipf distributions with their unique ranks of files and Zipf exponents. At each change point, these parameters vary randomly. The change points and Zipf parameters are all unknown to the caching agents. We only limit the Zipf exponent $\beta$ to be in the range $[1.1, 1.5]$. Then we plot the average of the percentages of the average transmission delay reduction over time as $\overline{\eta}_T = \frac{1}{T} \sum\limits_{t = 1}^{T} \eta_t$, for $t = 1, 2, ..., 40000$. As shown in Fig. \ref{fig:timeVarying}, the proposed framework achieves a lower performance at the beginning, because unlike the other three caching policies, the proposed framework doesn't directly collect the statistics from the users' requests, but generally adjust the parameters of the neural networks and learn the popularity patterns of the files. After the neural networks trained themselves well, the proposed framework achieves the best long-term performance. And at each time the popularity distribution changes, even though the performance slightly drops as the actor-critic framework updates the parameters to adapt to the new pattern, it is able to reach back to the previous level within a reasonable time frame because the previous experience has trained the network well. The LFU policy performs the best at the beginning, but due to the frequency pollution, the performance drops quickly at the first change point and goes all the way down. For the LRU and FIFO policies, the performances are stable, because the cache size is limited and the files that are used to be popular and less popular after the change can be replaced in a relatively short amount of time. However, as evidenced in this figure, the proposed framework is more suitable be to applied in scenarios that require long-term high performance and stability.

\section{Conclusion}
In this work, we have considered edge caching in cellular networks. To solve the dynamic problem of how to make caching decisions based on a partially observable Markov decision process, we have proposed a multi-agent framework based on deep actor-critic reinforcement learning, which enables each base station to make its own caching decision while competing for chances to transmit and also cooperating to achieve minimum overall transmission delay. We have designed the multi-agent actor-critic content caching algorithm. We have analyzed the performance of the proposed framework, and we have provided comparisons among the performances achieved by the proposed framework and three other caching algorithms, namely, LRU, LFU, and FIFO. The performances have been evaluated in terms of the overall average transmission delay reduction. Through simulation results, superior performance of the proposed framework has been demonstrated under different conditions in the experiments, and its excellent ability to adapt to changing environments is highlighted.

\bibliographystyle{ieeetr}
\bibliography{MultiAgent}

\end{document}